\documentclass[%
prl,
twocolumn,
superscriptaddress,
 amsmath,amssymb,
]{revtex4-1}

\usepackage{graphicx}
\usepackage{dcolumn}
\usepackage{bm}
\usepackage{siunitx}
\usepackage{xcolor}

\begin{document}

\title{Structural and magnetic investigation of the interfaces of Fe\textsubscript{3}O\textsubscript{4}/MgO(001) with and without NiO interlayer}
\author{\underline{Tobias Pohlmann}}
\email{tobias.pohlmann@desy.de}
\affiliation{Deutsches Elektronen-Synchrotron DESY, Notkestr. 85, 22607 Hamburg, Germany}
\affiliation{Department of Physics, Osnabr\"uck University, 49076 Osnabr\"uck, Germany}
\author{Florian Bertram}
\affiliation{Deutsches Elektronen-Synchrotron DESY, Notkestr. 85, 22607 Hamburg, Germany}
\author{Jannis Thien}
\affiliation{Department of Physics, Osnabr\"uck University, 49076 Osnabr\"uck, Germany}
\author{Jari Rodewald}
\affiliation{Department of Physics, Osnabr\"uck University, 49076 Osnabr\"uck, Germany}
\author{Kevin Ruwisch}
\affiliation{Department of Physics, Osnabr\"uck University, 49076 Osnabr\"uck, Germany}
\author{Timo Kuschel}
\affiliation{Center for Spinelectronic Materials and Devices, Department of Physics, Bielefeld University, Universit\"atsstr. 25, 33615 Bielefeld, Germany}
\author{Eugen Weschke}
\affiliation{Helmholtz-Zentrum Berlin f\"ur Materialien und Energie, Wilhelm-Conrad-R\"ontgen-Campus BESSY II, Albert-Einstein-Strasse 15, 12489 Berlin, Germany}
\author{Karsten K\"upper}
\affiliation{Department of Physics, Osnabr\"uck University, 49076 Osnabr\"uck, Germany}
\author{Joachim Wollschl\"ager}
\affiliation{Department of Physics, Osnabr\"uck University, 49076 Osnabr\"uck, Germany}
\date{\today}
             
\begin{abstract}
We present an investigation on the structural and magnetic properties of the interfaces of $\mathrm{Fe_3O_4/MgO(001)}$ and $\mathrm{Fe_3O_4/NiO/MgO(001)}$ by extracting cation-selective magnetooptical depth profiles by means of x-ray magnetic reflectivity (XRMR) in combination with charge-transfer multiplet simulations of x-ray magnetic circular dichroism (XMCD) data.
For $\mathrm{Fe_3O_4/MgO(001)}$, the magnetooptical depth profiles at the $\mathrm{Fe^{2+}_{oct}}$ and the $\mathrm{Fe^{3+}_{oct}}$ resonant energies follow exactly the structural profile, while the magnetooptical depth profile at the $\mathrm{Fe^{3+}_{tet}}$ resonance is offset by $3.2\pm1.3\,$\r{A} from the interface, consistent with a B-site interface termination of $\mathrm{Fe_3O_4}$ with fully intact magnetic order. In contrast, for $\mathrm{Fe_3O_4/NiO(001)}$, the magnetooptical depth profiles at the $\mathrm{Fe^{2+}_{oct}}$ and the $\mathrm{Ni^{2+}}$ resonances agree with the structural profile, but the interface positions of the magnetooptical depth profiles at the $\mathrm{Fe^{3+}_{oct}}$ and the $\mathrm{Fe^{3+}_{tet}}$ resonances are laterally shifted by $3.3\pm 1.4\,$\r{A} and $2.7\pm0.9\,$\r{A}, respectively, not consistent with a magnetically ordered stoichiometric interface. This may be related to an intermixed $\mathrm{(Ni,Fe)O}$ layer at the interface. The magnetooptical depth profiles at the Ni $L_3$ edge reveal uncompensated magnetic moments throughout the NiO film.
\end{abstract}
\pacs{Valid PACS appear here}
\maketitle


\section{\label{sec:level1}Introduction}
Magnetite ($\mathrm{Fe_3O_4}$) is a half-metallic ferrimagnet in the inverse spinel structure. This structure consists of a cubic close-packed oxygen lattice whose interstitial sites are populated by three different iron species: 
1/2 of the octahedral B-sites are occupied randomly by divalent $\mathrm{Fe^{2+}_{oct}}$ and trivalent $\mathrm{Fe^{3+}_{oct}}$ cations, and 1/8 of the tetrahedral A-sites are occupied by trivalent $\mathrm{Fe^{3+}_{tet}}$ cations. The two octahedrally coordinated species $\mathrm{Fe^{2+}_{oct}}$ and $\mathrm{Fe^{3+}_{oct}}$ are ferromagnetically coupled by double exchange, while the $\mathrm{Fe^{3+}_{oct}}$ and $\mathrm{Fe^{3+}_{tet}}$ cations couple antiferromagnetically via superexchange. Therefore, the magnetic moments of the $\mathrm{Fe^{3+}}$ cations compensate each other and the resulting macroscopic moment of $\mathrm{Fe_3O_4}$ of $4.07\,\mathrm{\frac{\mu_B}{f.u.}}$ is determined by the magnetic moments of the $\mathrm{Fe^{2+}_{oct}}$ cations \cite{Weiss29}.

Due to these magnetic properties, magnetite is a long-standing candidate to contribute to all-oxide thin-film spintronic devices, as a source for spin-polarized currents \cite{Moussy13,Coey03,Bibes07,Zutic04,Moyer15,Marnitz15}. These kinds of devices utilize the fact that many metal oxides with varying electronic and magnetic properties grow in spinel or rock-salt structures, such as the conducting ferrimagnets $\mathrm{Fe_3O_4}$ and $\gamma$-$\mathrm{Fe_2O_3}$, the insulating ferrimagnets $\mathrm{NiFe_2O_4}$ and $\mathrm{CoFe_2O_4}$, the insulating antiferromagnets NiO, CoO and FeO, or the insulating diamagnets MgO and $\mathrm{MgAl_2O_4}$, which all share a cubic close-packed oxygen lattice with very similar lattice constants \cite{Moussy13}. This allows epitaxial growth of film stacks with a large variety of spin electronic functionality but with little strain and thus supposedly well-matching interfaces.

However, a drawback of this concept is that the structural similarity of these metal oxides also means that undesired modifications at their interfaces are difficult to detect, such as interdiffusion of Mg \cite{Kim09}, Ni \cite{Farrow00,Kuschel16} or Co \cite{Rodewald19} into $\mathrm{Fe_3O_4}$ films, or the transformation of the different iron oxides into each other \cite{Bertram11, Bertram12}. 
All-oxide spintronic devices with $\mathrm{Fe_3O_4}$ electrodes did indeed not prove to be very successful yet; their shortcomings were speculated to stem from magnetic dead layers at the substrate interface \cite{VanDerHeijden96,Zaag00} or other interface effects \cite{Marnitz15}.

Of particular interest have been the interfaces between Fe\textsubscript{3}O\textsubscript{4} films and the tunnel barrier material MgO \cite{Marnitz15,Zaag00,Kado08}, as well as the interface between Fe\textsubscript{3}O\textsubscript{4} and antiferromagnetic films, e.g. NiO, exhibiting exchange bias \cite{Keller02,Gatel05,Krug08,Kuepper16}. This effect can cause a shift of the coercive fields of the ferrimagnetic Fe\textsubscript{3}O\textsubscript{4} film and be used to pin its magnetization state. 

In this study, we investigate the structural and magnetic properties of the $\mathrm{Fe_3O_4/NiO}$ and $\mathrm{Fe_3O_4/MgO}$ interfaces. We grow $\mathrm{Fe_3O_4}$ single layers and $\mathrm{Fe_3O_4/NiO}$ bilayers on MgO(001) by reactive molecular beam epitaxy (RMBE) and investigate the distribution and magnetic order of the three cations $\mathrm{Fe^{2+}_{oct}}$, $\mathrm{Fe^{3+}_{tet}}$ and $\mathrm{Fe^{3+}_{oct}}$ of $\mathrm{Fe_3O_4}$ and of the $\mathrm{Ni^{2+}}$ cations of NiO by x-ray resonant magnetic reflectivity (XRMR) combined with charge-transfer multiplet analysis of x-ray magnetic circular dichroism (XMCD) spectra. We find that the magnetic structure of $\mathrm{Fe_3O_4}$ on MgO is intact down to the interface, while on the $\mathrm{Fe_3O_4/NiO}$ interface the data indicates a disturbed magnetic order. XRMR data on the Ni $L_3$ edge indicate uncompensated magnetic moments in the antiferromagnet NiO. 
\section{\label{sec:level2}Experimental details}
The deposition and characterization methods of the samples followed the ones presented in Refs. \cite{Kuepper16,Kuschel16}.
We prepared $\mathrm{Fe_3O_4/MgO(001)}$ and $\mathrm{Fe_3O_4/NiO/MgO(001)}$ samples in a multichamber ultra-high-vacuum system with a base pressure of $p_0 < 1 \times 10^{-8}\,$mbar.
Before deposition, the MgO(001) substrates were annealed at $400^\circ$C in an oxygen atmosphere of $1 \times 10^{-4}\,$mbar for 1 hour. Our films were grown by RMBE. 
For the NiO, we deposited nickel in an oxygen pressure of $1 \times 10^{-5}\,$mbar, and for the $\mathrm{Fe_3O_4}$, we deposited iron in an oxygen pressure of $5 \times 10^{-6}\,$mbar.
We limited the substrate temperature to $250^\circ$C in order to avoid interdiffusion of Mg into the films \cite{Kim09}. After growth, the electronic structure of the samples were characterized \textit{in~situ} by x-ray photoelectron spectroscopy (XPS) using a Phoibos HSA 150 hemispherical analyzer and an Al K$\alpha$ anode, and their surface structure by low-energy electron diffraction (LEED). The Fe $2p$ XPS spectra show the $\mathrm{Fe^{2+}}$ and the $\mathrm{Fe^{3+}}$ features typical for $\mathrm{Fe_3O_4}$, and the LEED patterns confirm the characteristic ($\sqrt{2}\times \sqrt{2}$)R$45^\circ$ surface structure of $\mathrm{Fe_3O_4}$ \cite{Mariotto02,Bliem14} (both not shown here).

$\mathrm{Fe_3O_4/MgO(001)}$ and $\mathrm{Fe_3O_4/NiO/MgO(001)}$ samples were transported under ambient conditions to BESSY II for x-ray absorption spectroscopy (XAS), XMCD, x-ray reflectivity (XRR) and XRMR on the XUV diffractometer at beamline $\text{UE46\_PGM-1}$ \cite{UE46PGM1}. The samples were placed between two permanent magnets in a magnetic field of $200\,$mT at room temperature. The x-rays had a degree of 90\% circular polarization. 

\begin{figure}[t]
\includegraphics[scale=0.4]{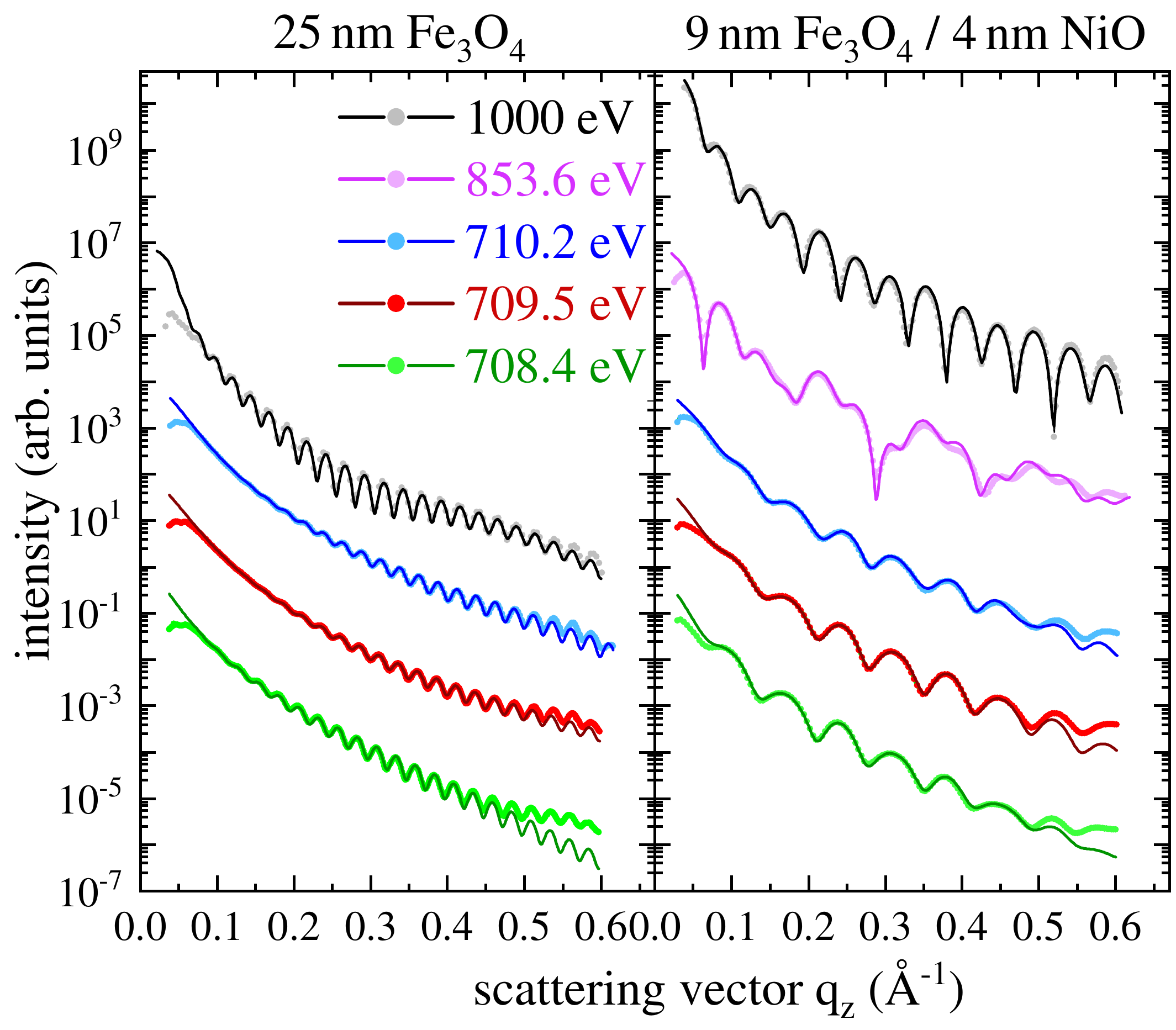}
\caption{\label{p3fig_xrr} XRR measurements of the two samples recorded at the three resonant energies of the Fe \textit{L}\textsubscript{3} XMCD spectrum of Fe\textsubscript{3}O\textsubscript{4} (708.4$\,$eV, 709.5$\,$eV, 710.2$\,$eV), the energy of the Ni \textit{L}\textsubscript{3} maximum (853.6$\,$eV) and at an off-resonant energy (1000$\,$eV).}
\end{figure}

All XAS and XMCD spectra were recorded in total electron yield (TEY) mode with an incidence glancing angle of $30^\circ$. XRR and XRMR curves were obtained by $\theta$-$2\theta$ scans in the range $2\theta=0^\circ - 140^\circ$ at selected resonant photon energies with both right and left circularly polarized x-rays.
The structural properties of the samples (thickness $d$, roughness $\sigma$) obtained by XRR  at an off-resonant energy ($1000\,$eV, cf. Fig.~\ref{p3fig_xrr}) are summarized in Tab.~\ref{p3tab_thicc}.  

In order to obtain magnetic information with higher depth sensitivity, we measured XAS and XMCD in total fluorescence yield mode (TFY) and reflection mode on a different $\mathrm{Fe_3O_4/NiO/MgO}$ sample at Diamond Light Source (DLS), on the RASOR diffractometer of beamline I10. The sample was placed in a similar magnet setup, again in a magnetic field of $200\,$mT and at room temperature. Here, the x-rays had a degree of circular polarization of 99\%. The structural properties of this sample can be found in Tab.~\ref{p3tab_thicc}, too.

\begin{table}[t]
\begin{minipage}{0.07\textwidth}
~
\end{minipage}
\hfill
\begin{minipage}{0.2\textwidth}
\begin{center}
BESSY II UE46$\_$PGM-1
\end{center}
\end{minipage}
\hfill
\begin{minipage}{0.15\textwidth}
DLS I10
\end{minipage}
\centering
\begin{tabular}{l|cc|c}
~ & $\mathrm{Fe_3O_4/MgO}$ & $\mathrm{Fe_3O_4/NiO/MgO}$ & $\mathrm{Fe_3O_4/NiO/MgO}$ \\ 
\hline 
$d_\mathrm{Fe_3O_4}$ & $25.2\pm0.3\,$nm & $9.3\pm0.1\,$nm & $17.6\pm0.1\,$nm \\ 
$d_\mathrm{NiO}$ & -- & $4.3\pm0.1\,$nm & $27.3\pm0.2\,$nm \\  
\hline
$\sigma_\mathrm{Fe_3O_4}$ & $3.3\pm0.5\,$\r{A} & $3.2\pm0.5\,$\r{A} & $2.0\pm0.6\,$\r{A} \\ 
$\sigma_\mathrm{NiO}$& -- & $3.0\pm0.2\,$\r{A} & $4.2\pm0.8\,$\r{A}\\ 
$\sigma_\mathrm{MgO}$& $3.5\pm0.5\,$\r{A} & $2.7\pm0.5\,$\r{A} & $2.5\pm0.3\,$\r{A}\\ 
\end{tabular} 
\caption{\label{p3tab_thicc} Film thicknesses $d_i$ and rms roughnesses $\sigma_i$ of the three investigated samples, obtained from off-resonant XRR measurements recorded at 1000$\,$eV. Corresponding data and fits for the UE46$\_$PGM-1 samples are shown in Fig. \ref{p3fig_xrr}.}
\end{table}

\section{Data analysis}
\subsection{XMCD}
\begin{figure}[t]
\centering
\includegraphics[scale=0.4]{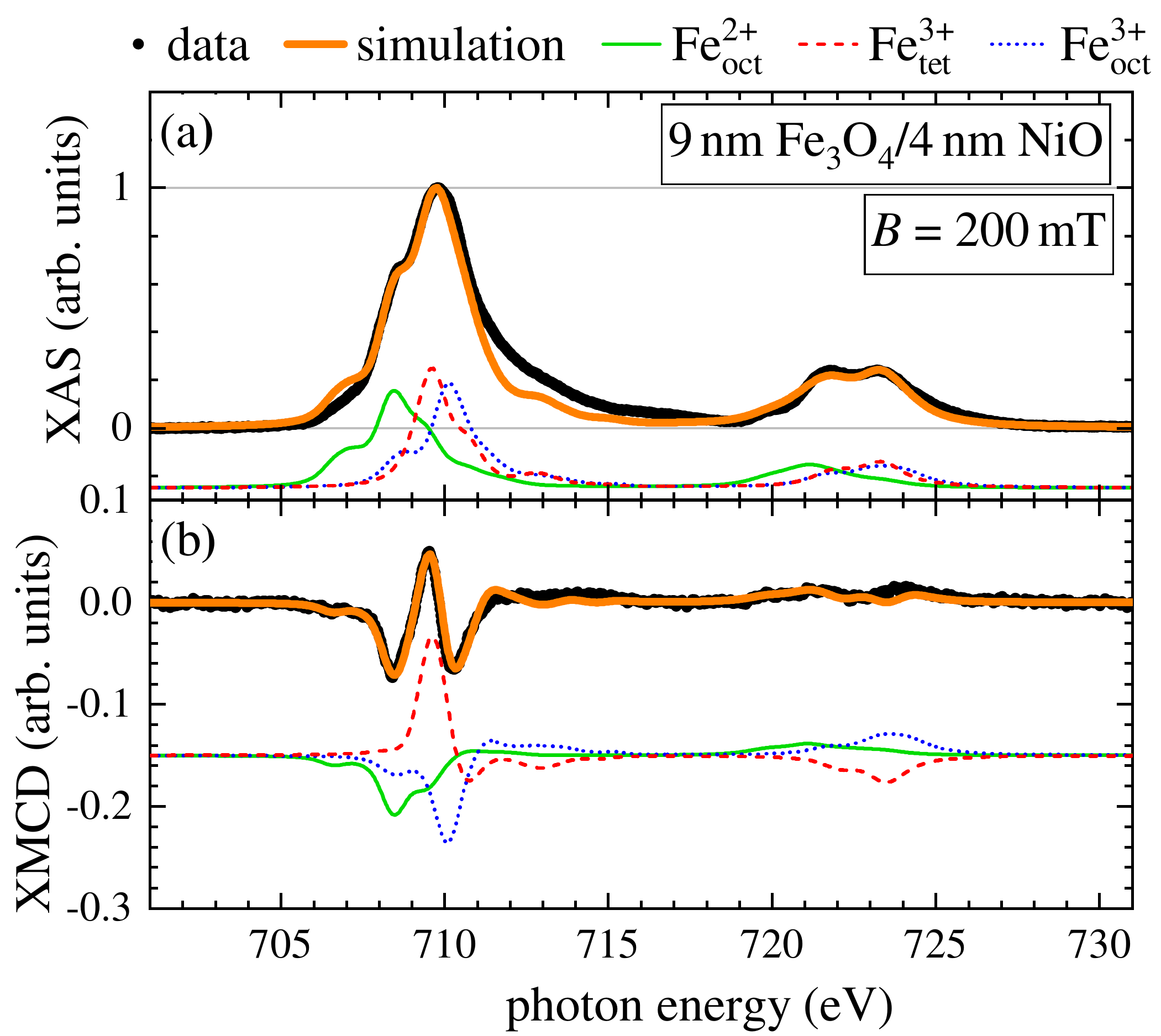}
\caption{\label{p3fig_XMCD}(a) XAS and (b) XMCD spectrum at the Fe \textit{L}\textsubscript{2,3} edges for the Fe\textsubscript{3}O\textsubscript{4}/NiO/MgO sample, taken at 200$\,$mT external magnetic field, at room temperature and in TEY mode. A step function was subtracted from the XAS spectrum. Black circles are data points; green, red and blue spectra are multiplet calculations for the three cation species of Fe\textsubscript{3}O\textsubscript{4} assuming an ideal 1:1:1 stoichiometry; the orange line is their sum. The cation spectra are offset for better visibility.}
\end{figure}

To obtain cation- and lattice-site-selective magnetooptical depth profiles, first the XMCD spectra have to be analyzed. Figures \ref{p3fig_XMCD}(a) and (b) show exemplarily XAS and XMCD spectra, respectively, of the Fe $L_{2,3}$ edges of the $\mathrm{Fe_3O_4/NiO/MgO}$ sample. Charge-transfer multiplet calculations of the three Fe cations of $\mathrm{Fe_3O_4}$ using the Thole code \cite{VanDerLaan97} with assistance of CTM4XAS \cite{deGroot05,Stavitski10} provide the three individual XAS and XMCD spectra shown below the data. For these calculations, we used Ref. \cite{Kuepper16} as a starting point: we assumed the three-cation model with crystal field energies of \mbox{$\mathrm{10Dq^{oct}} = 1.0\,$eV} in octahedral and \mbox{$\mathrm{10Dq^{tet}} = -0.6\,$eV} in tetrahedral coordination. The splittings between the initial and final charge-transfer states were chosen as \mbox{$\Delta_\mathrm{init} = 6\,$eV} and \mbox{$\Delta _\mathrm{final}= 5\,$eV}, and for the exchange splitting, \mbox{$g \cdot \mu_\mathrm{B}=12\pm 1\,$meV} was used.
The multiplet states resulting from these calculations were compared to the experimental data by assuming a Gaussian instrumental broadening of $0.25\,$eV and a Lorentzian lifetime broadening of $0.3\,$eV at $L_3$ and $0.6\,$eV at $L_2$. 
Adding the three individual cation spectra with a 1:1:1 ratio, as expected for Fe\textsubscript{3}O\textsubscript{4}, results in a total XAS and a total XMCD spectrum (orange lines in Figs.~\ref{p3fig_XMCD}(a),(b), respectively), which fit both the XAS and the XMCD data well. The multiplet analysis reveals that at those energies for which the XMCD spectrum has its extrema ($708.4\,$eV, $709.5\,$eV, $710.2\,$eV), most of the XMCD signal originates from one dominant cation species \cite{Tobi2020}. The individual contributions of each cation species to the XMCD spectrum at these three energies can be found in Tab. \ref{p3tab_resonances}. Therefore, XRMR measurements on those energies are mostly sensitive to one specific cation species. This allows to disentangle the contributions of the individual  cations to the magnetooptical depth profiles.

\begin{table}[b]
\begin{minipage}{0.5\textwidth}
\begin{tabular}{cccc}
Energy & $\mathrm{Fe^{2+} _{oct}}$ & $\mathrm{Fe^{3+} _{tet}}$ & $\mathrm{Fe^{3+} _{oct}}$ \\ 
\hline
$708.4\,$eV & $70 \pm 5\%$ & $-8 \pm 3\%$ & $22 \pm 5\%$  \\ 
$709.5\,$eV & $19 \pm 3\%$ & $-63 \pm 3\%$ & $18\pm 3\%$ \\ 
$710.2\,$eV & $4\pm 2\%$ & $-25 \pm 8 \%$ & $71 \pm 10 \%$ \\ 
\end{tabular}
\end{minipage}
\hfill
\begin{minipage}{0.49\textwidth}
\caption{\label{p3tab_resonances} Contributions of the three cation species to the extrema in the XMCD spectrum in Fig. \ref{p3fig_XMCD}(b) as obtained by the multiplet analysis.}
\end{minipage} 
\end{table}

\begin{figure}[t]
\centering
\includegraphics[scale=0.4]{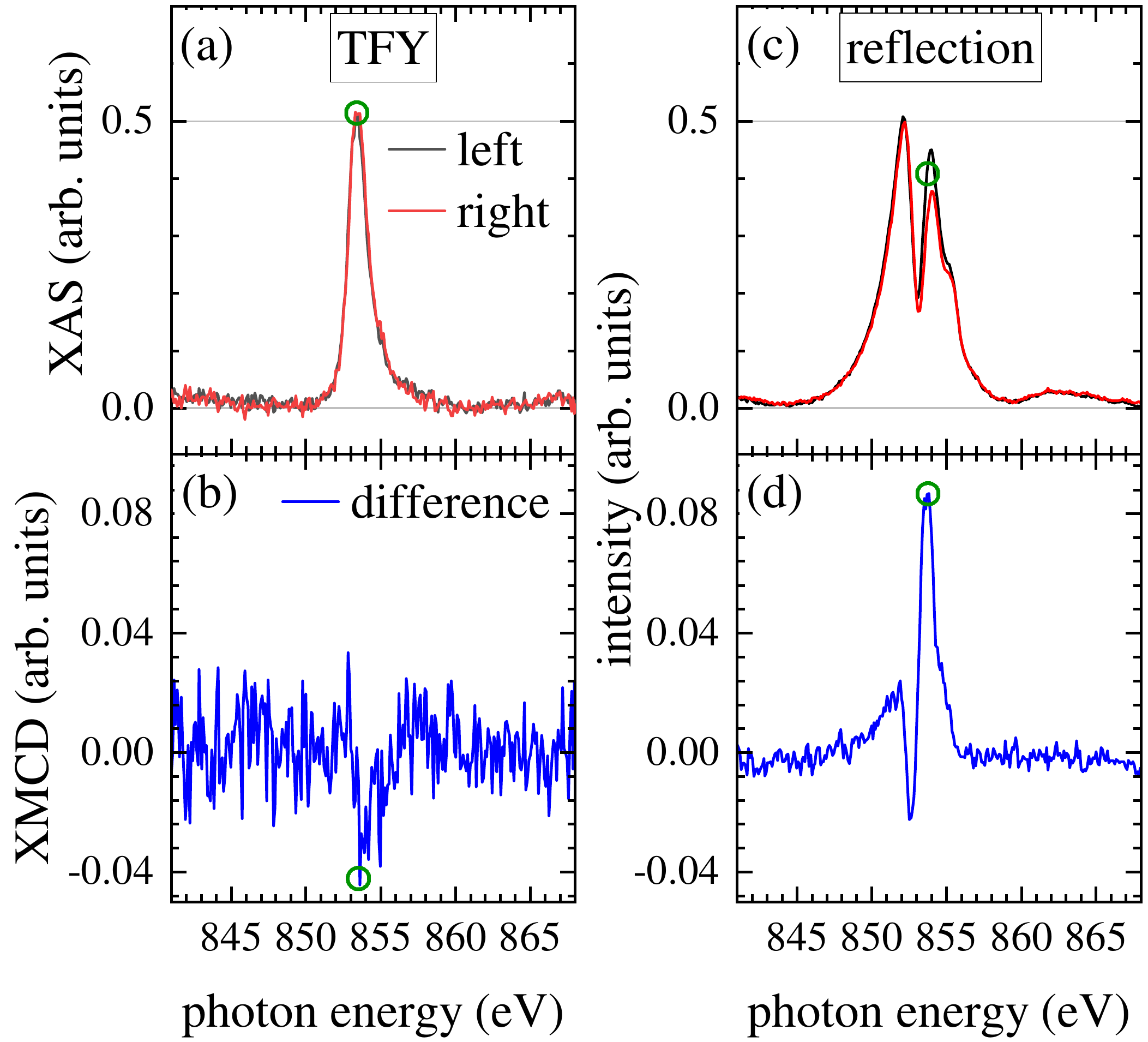}
\caption{\label{p3fig_NiMCD} (a) XAS and (b) XMCD spectra in TFY mode at the Ni \textit{L}\textsubscript{3} edge for a 
18$\,$nm~Fe\textsubscript{3}O\textsubscript{4}/27$\,$nm~NiO/MgO sample. (c) Energy scans of the reflected intensity for left and right circularly polarized x-rays. (d) Difference of the curves in (c). The green circles indicate the energy at which the XRMR curves were taken (853.6$\,$eV). The XAS data are normalized to the maximum of the XAS spectrum XAS\textsuperscript{left}+XAS\textsuperscript{right}. All data are recorded at a fixed incidence glancing angle of 30$^\circ$.}
\end{figure}

Since the TEY mode has a probing depth of about $3\,$nm in $\mathrm{Fe_3O_4}$ \cite{Gomes14}, the TEY signal from the buried NiO film was strongly attenuated at the Ni $L_3$ edge. For a clearer signal, we brought a $\mathrm{18\,nm~Fe_3O_4/27\,nm~NiO/MgO}$ sample to beamline I10 of DLS, and measured XAS and XMCD in TFY mode at a fixed incident angle of $30^\circ$, whose probing depth is only limited by the x-ray attenuation length in $\mathrm{Fe_3O_4}$ of about $80\,$nm \cite{Henke93} (cf. Figs. \ref{p3fig_NiMCD}(a),(b)). Simultaneously, we measured the reflected intensity in order to obtain an estimate of the magnetooptic effects in reflection (cf. Figs. \ref{p3fig_NiMCD}(c),(d)). 

\subsection{XRMR}
The XRMR data were recorded by measuring XRR curves at resonant photon energies $E_i$ with extrema in the XMCD signal (maximum at $708.4\,$eV, minimum at $709.5\,$eV, maximum at  $710.2\,$eV, see Fig. \ref{p3fig_XMCD}(b)) with both left and right circularly polarized x-rays. 
Resonant 'non-dichroic' XRR curves were obtained by averaging the signals $I^\sigma$ from both helicities ($\sigma=$right/left)
\begin{equation}
I = (I^\mathrm{right}+I^\mathrm{left})/2
\end{equation}
and the XRMR asymmetry ratios by subtracting and normalizing them:
\begin{equation}
\Delta I = \frac{I^\mathrm{right}-I^\mathrm{left}}{I^\mathrm{right}+I^\mathrm{left}}.
\label{p3eq_asymmetry}
\end{equation}
These curves were then fitted with the Zak matrix formalism using the software ReMagX \cite{Macke14} to determine the depth profiles of the complex refractive index $n(z)$
\begin{equation}
n(z) = 1-\delta(z)+ i\beta(z)
\end{equation} 
along the film height $z$.
The optical dispersion $\delta$ and the optical absorption $\beta$ can be split into non-magnetic components $\delta_0$, $\beta_0$ and magnetooptical components $\Delta \delta$, $\Delta \beta$. In the case of an in-plane magnetic field longitudinal to the x-ray beam as applied here, they can be written as \cite{Macke14}
\begin{eqnarray}
\delta(z) = \delta _0(z) \mp \Delta \delta (z) \cdot \mathrm{cos}(\theta)\\ \beta(z) = \beta _0(z) \pm \Delta \beta(z)\cdot \mathrm{cos}(\theta) \ \ \ ,
\end{eqnarray}
for which the magnetooptical contributions depend on the x-ray incidence glancing angle $\theta$ and their sign on the helicity of the x-rays.
The optical absorption $\beta_0$ is proportional to the XAS signal, while the magnetooptical absorption $\Delta \beta$ is proportional to the XMCD signal. Thus, $\Delta \beta(z)$ is a measure of the magnetization along the film depth. 
A detailed review of the XRMR method and the software is given in Ref. \cite{Macke14}, and a conclusive recipe for fitting XRMR data can be found in Refs. \cite{Kuschel15,Klewe16,Krieft20}.  

\section{\label{sec:level3}Results}
\begin{figure}[t]
\includegraphics[scale=0.4]{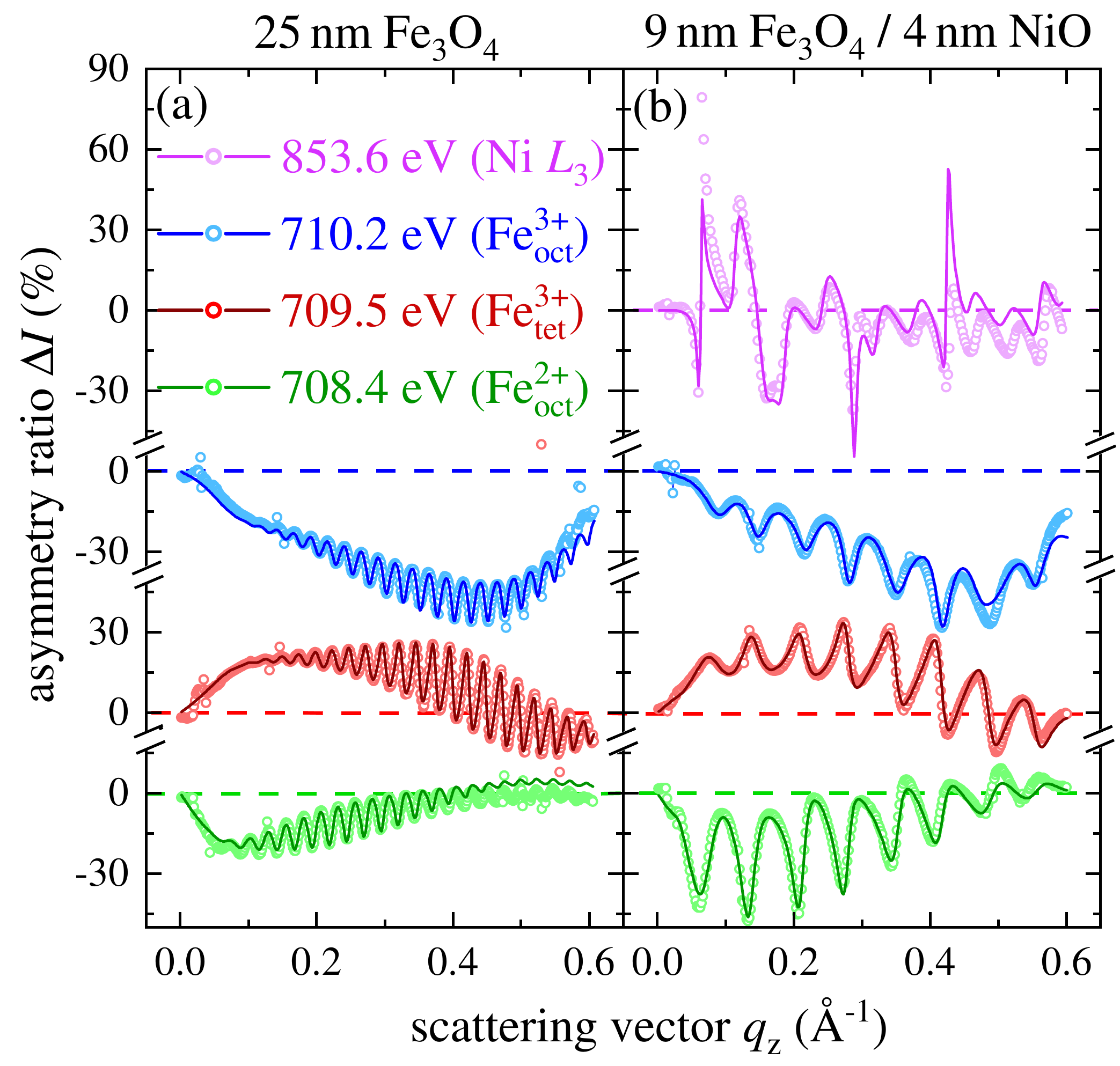}
\caption{\label{p3fig_XRMR} XRMR data (open circles) and fits (solid lines) from (a) the Fe\textsubscript{3}O\textsubscript{4}/MgO and (b) the Fe\textsubscript{3}O\textsubscript{4}/NiO/MgO sample, recorded at the three resonant energies of the Fe XMCD \textit{L}\textsubscript{3} edge and the Ni \textit{L}\textsubscript{3} edge. The magnetooptical depth profile models corresponding to the fits in (a) are displayed in Fig. \ref{p3fig_expProf}(b), and the interface regions of the magnetooptical depth profiles corresponding to the fits in (b) are shown in Fig. \ref{p3fig_expProf}(c).}
\end{figure} 
Figure \ref{p3fig_NiMCD}(a) shows the XAS spectra at the Ni $L_3$ edge of the $\mathrm{18\,nm~Fe_3O_4/27\,nm~NiO/MgO}$ sample recorded with left and right circularly polarized x-rays in TFY mode, and Fig. \ref{p3fig_NiMCD}(b) shows their difference. The XMCD signal is about 3\% of the XAS maximum. In reflection, the magnetooptical effects become even more apparent. Note that the data are recorded at fixed incidence angle of $\theta=30^\circ$. Therefore the scattering vector $q_z$ changes from $0.426\,\mathrm{\r{A}^{-1}}$ to $0.440\,\mathrm{\r{A}^{-1}}$ along with the energy in the range displayed in Figs.~\ref{p3fig_NiMCD}(c),(d). At the Ni $L_3$ resonant energy of $853.6\,$eV, the magnetooptically induced change of the reflected intensity ranges up to about 8\% demonstrating a strong magnetooptical signal from NiO. This energy was chosen for the XRMR measurements at the Ni $L_3$ resonance and is indicated by the green circles in Fig. \ref{p3fig_NiMCD}.

\begin{figure}[t]
\centering
\includegraphics[scale=0.4]{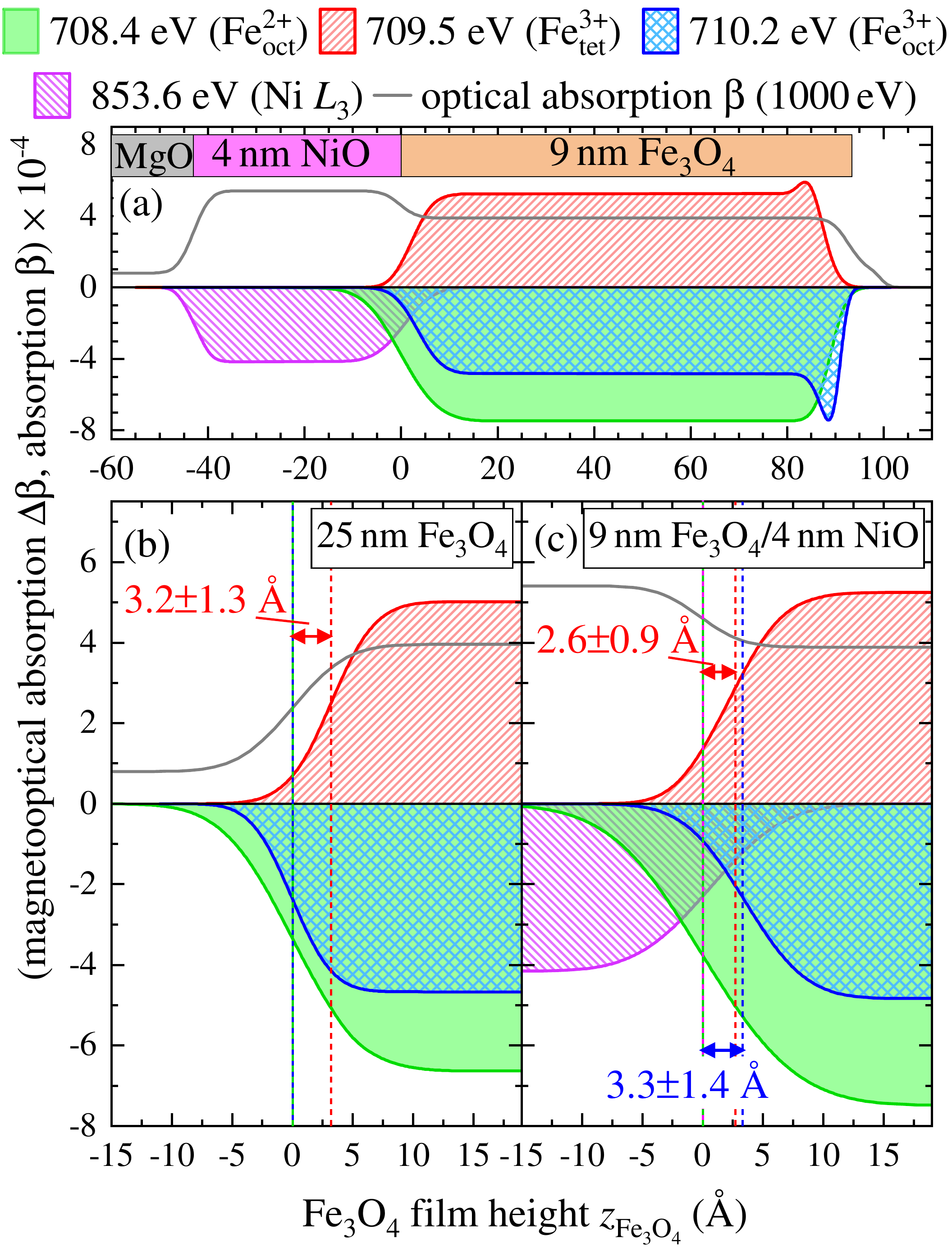}
\caption{\label{p3fig_expProf} (a) $\Delta \beta$(\textit{z}) depth profiles at the three Fe \textit{L}\textsubscript{3} resonant energies and the Ni \textit{L}\textsubscript{3} energy of the
Fe\textsubscript{3}O\textsubscript{4}/NiO/MgO sample, corresponding to the fits in Fig. \ref{p3fig_XRMR}(b). (b) Close-up of the interface region of the Fe\textsubscript{3}O\textsubscript{4}/MgO film, obtained from the fits in Fig. \ref{p3fig_XRMR}(a).  (c) Close-up of the Fe\textsubscript{3}O\textsubscript{4}/NiO interface region of the magnetooptical depth profiles in (a). 
The grey lines indicate the optical absorption profiles $\beta _\text{off-res}$(\textit{z}) obtained by off-resonant XRR fits. Dashed lines highlight the interface positions of the magnetooptical depth profiles.}
\end{figure}

The data and fits of the resonant XRR measurements can be found in Fig. \ref{p3fig_xrr} alongside with the off-resonant XRR curves recorded with $1\,$keV photons.
Figures \ref{p3fig_XRMR}(a) and (b) show the XRMR data for the $\mathrm{Fe_3O_4/MgO}$ and the $\mathrm{Fe_3O_4/NiO/MgO}$ samples at the three Fe $L_3$ resonant energies $708.4\,$eV, $709.5\,$eV and $710.2\,$eV, and the Ni $L_3$ resonant energy $853.6\,$eV, together with their respective fits, which describe the data very well. 
The fits in Fig.~\ref{p3fig_XRMR}(b) were obtained from the magnetooptical depth profile models which are displayed in Fig. \ref{p3fig_expProf}(a) for all four resonant energies of the $\mathrm{Fe_3O_4/NiO/MgO}$ sample. The grey line represents the optical absorption $\beta_\text{off-res}$ obtained from the off-resonant XRR measurement, and represents the structural depth profiles of the sample. It has three plateaus corresponding to the MgO substrate, the NiO film and finally the $\mathrm{Fe_3O_4}$ film, as illustrated by the sketch on top of the panel.
The filled areas are the magnetooptical depth profiles $\Delta \beta (z)$ at the resonant energies of the three iron cation species and the Ni $L_3$ energy, obtained from the XRMR asymmetry ratios. 

One notable feature of these magnetooptical depth profiles is found at the surface of the $\mathrm{Fe_3O_4}$ films. Here, a thin layer of enhanced magnetooptical absorption is observed at $709.5\,$eV and $710.2\,$eV, both for the $\mathrm{Fe_3O_4/MgO(001)}$ and the $\mathrm{Fe_3O_4/NiO/MgO(001)}$ samples. This is likely related to a modification of the cation stoichiometry at the $\mathrm{Fe_3O_4(001)}$ surface.
This effect is discussed in detail in Ref. \cite{Tobi2020} for $\mathrm{Fe_3O_4/MgO(001)}$, and it is interesting to note that it occurs on the $\mathrm{Fe_3O_4/NiO/MgO(001)}$ sample, too, as exemplified by Fig.~\ref{p3fig_expProf}(a). However, it is not the subject of the current study, which focuses on the $\mathrm{Fe_3O_4/MgO}$ and $\mathrm{Fe_3O_4/NiO}$ interfaces. 

Therefore, Fig. \ref{p3fig_expProf}(b) focuses upon the interface region of the magnetooptical depth profiles of the $\mathrm{Fe_3O_4/MgO}$ sample, according to the fits in Fig. \ref{p3fig_XRMR}(a). Both the interfaces of the magnetooptical depth profile at $708.4\,$eV (green) and of the one at $710.2\,$eV (blue) are collocated with the structural interface (grey line) at $z=0\,$\r{A}. However, the interface of the magnetooptical depth profile at $709.5\,$eV (red) is shifted by a distance $\Delta z_\mathrm{709.5\,eV}=3.2 \pm 1.3\,$\r{A} away from the interface into  the Fe\textsubscript{3}O\textsubscript{4} film. The roughnesses of magnetooptical depth profiles at both $\mathrm{Fe^{3+}}$ resonances follow the structural depth profile. In contrast, the roughness of the magnetooptical depth profile recorded at the $\text{Fe}^{2+}_\text{oct}$ energy, $\sigma_\mathrm{708.4\,eV}=4.4\pm 0.2\,$\r{A}, is slightly larger than the structural roughness $\sigma_\mathrm{substrate} = 3.5\pm 0.5\,$\r{A}. 

\begin{figure}[t]
\centering
\includegraphics[scale=0.4]{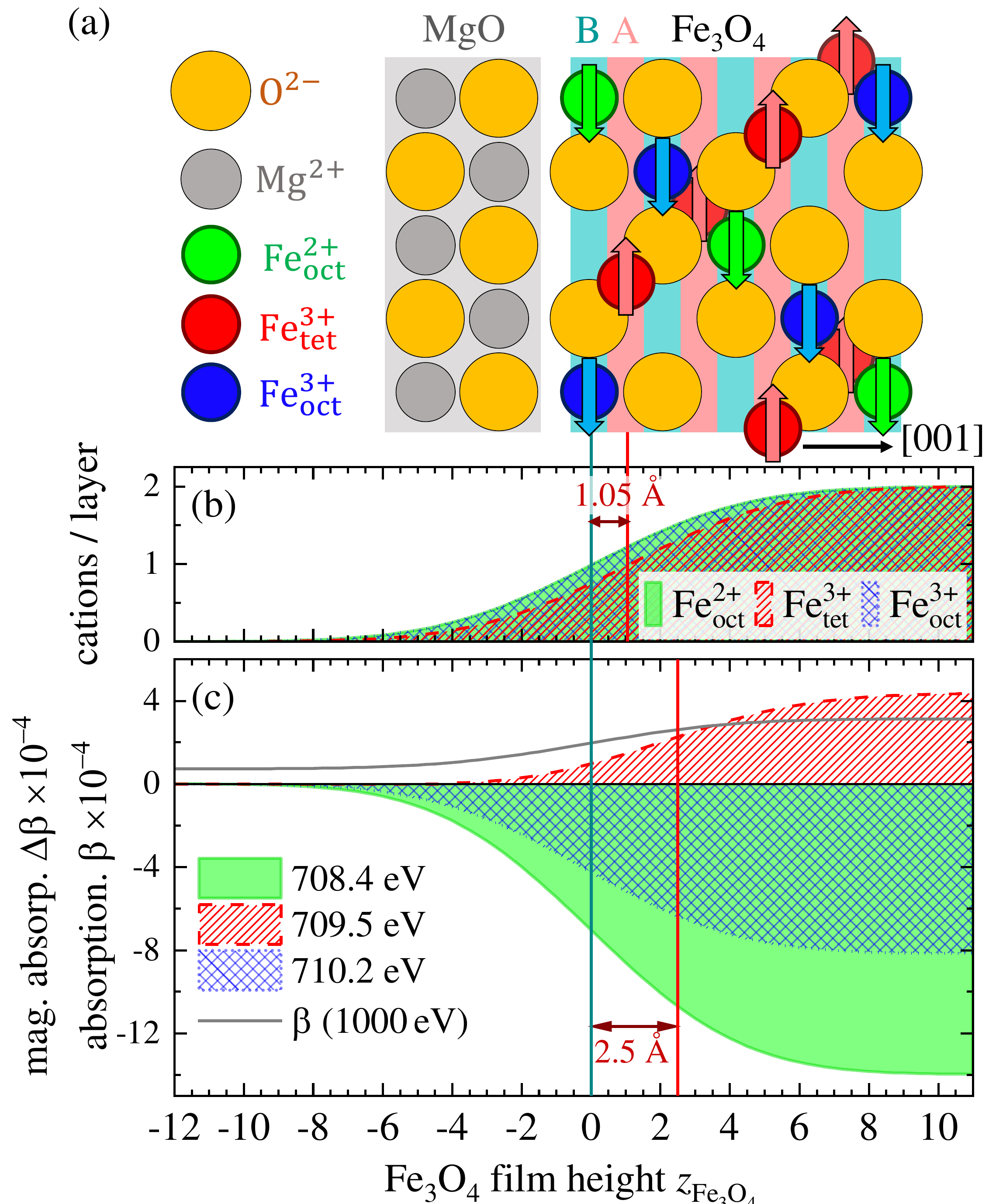}
\caption{\label{p3fig_simProf} (a) Illustration of the stacking order at a B-terminated Fe\textsubscript{3}O\textsubscript{4}/MgO(001) interface. The blue and red shaded areas indicate octahedral and tetrahedral layers, respectively. Arrows depict magnetic moments. (b) Simulated cation profiles at the interface of a Fe\textsubscript{3}O\textsubscript{4} film with a roughness of $\sigma$ = 3.5$\,$\r{A}, in scale with (a). (c) Expected magnetooptical depth profiles at the three resonant energies of the Fe \textit{L}\textsubscript{3} edge, stemming from the cation profiles in (b), and assuming the magnetooptical absorption contributions of the individual cation spectra in Fig. \ref{p3fig_XMCD}(b).}
\end{figure}

For the $\mathrm{Fe_3O_4/NiO/MgO}$ film, the results are slightly different. Figure \ref{p3fig_expProf}(c) shows the $\mathrm{Fe_3O_4/NiO}$ interface region of the $\mathrm{Fe_3O_4/NiO/MgO}$ sample. 
The XRMR data at the Ni $L_3$ edge can be well fitted with a homogeneous magnetization profile throughout the NiO film. The interfaces of the magnetooptical depth profiles at the Ni $L_3$ edge and at the $\mathrm{Fe^{2+}_{oct}}$-related resonance at $708.4\,$eV are collocated with the structural interface, indicating intact structural and magnetic order for both species. 
Notably, both their roughnesses are slightly higher, $\sigma _\mathrm{708.4\,eV}=6.3\pm0.7\,$\r{A}, $\sigma_\mathrm{853.6\,eV}=5.0\pm 0.7\,$\r{A}, compared to the structural roughness $\sigma _\mathrm{Fe_3O_4/NiO}=3.0\pm0.2\,$\r{A}.

In contrast to the magnetooptical depth profile at $708.4\,$eV, which directly follows the structural profile, the profiles at $709.5\,$eV and $710.2\,$eV are rising with shifts of $\Delta z_\mathrm{709.5\,eV}=2.6 \pm 0.9\,$\r{A} and $\Delta z_\mathrm{710.2\,eV}=3.3\pm 1.4 \,$\r{A}, respectively, apart from the interface, pointing to a lack of magnetooptical absorption at resonant energies of both $\mathrm{Fe^{3+}_{oct}}$ and $\mathrm{Fe^{3+}_{tet}}$ at the interface. 

\begin{figure*}
\centering
\includegraphics[scale=0.4]{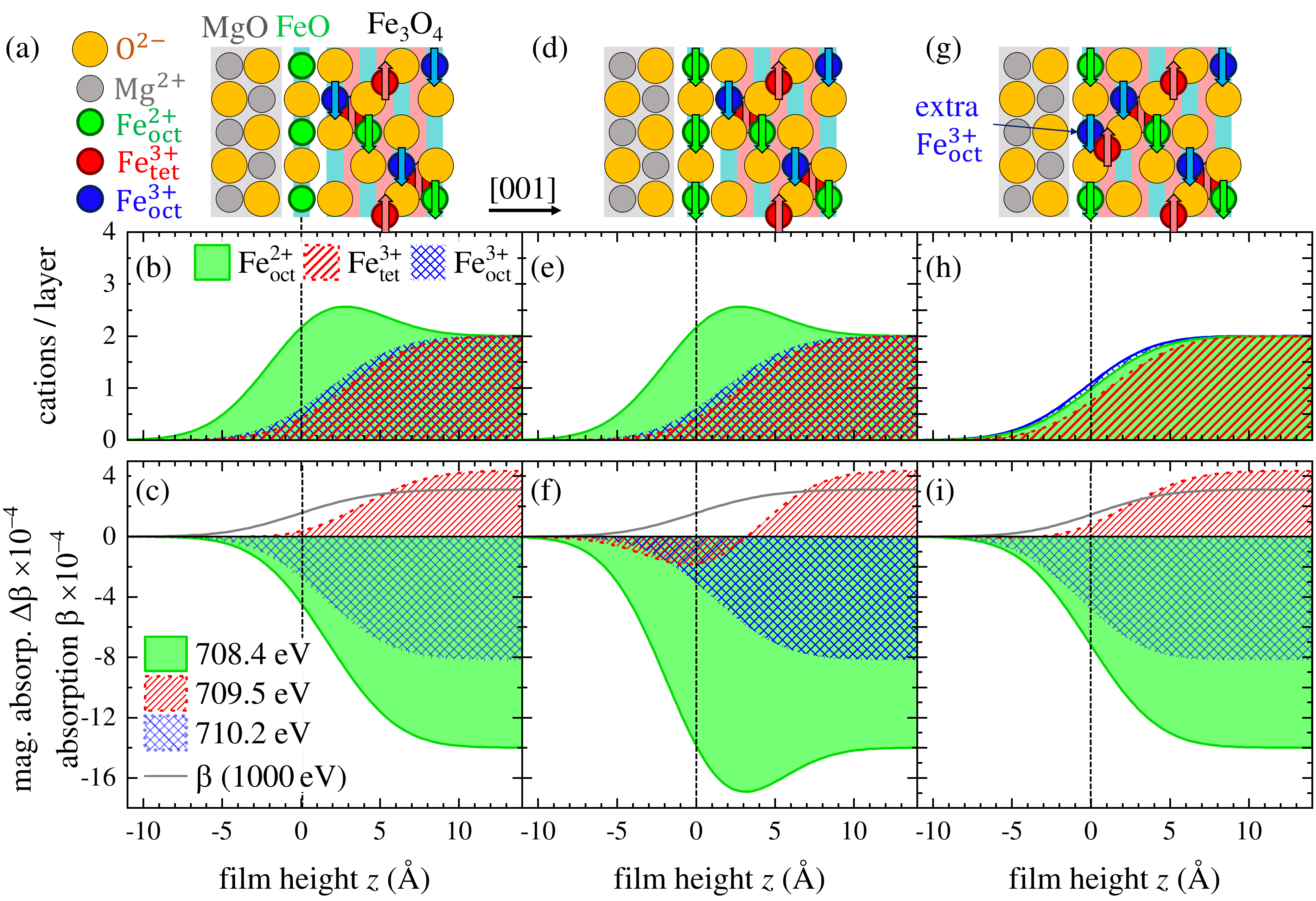}
\caption{\label{p3_figProfSim_extra} Simulated magnetooptical depth profiles for three models of the Fe\textsubscript{3}O\textsubscript{4}/MgO interface. (a),(d),(g) Illustrations of the stacking orders of Fe\textsubscript{3}O\textsubscript{4}/MgO interfaces (a) with a single non-magnetic Fe\textsubscript{1-$\delta$}O interlayer, (d) with a single magnetic Fe\textsubscript{1-$\delta$}O interlayer, (g) with an additional $\text{Fe}^{3+}_\text{oct}$ cation in the interface layer, following the model of Chang et al \cite{Chang16}. (b),(e),(h) Cation depth profiles corresponding to the illustrations in (a),(d),(g), respectively. (c),(f),(i) Magnetooptical profiles resulting from the cation profiles in (b),(e),(h), respectively. Distances of the rising edges of the magnetooptical profiles from the structural interface can be found in Tab.~\ref{p3_tabdistances}.} 
\end{figure*}

\begin{table*}
\begin{tabular}{c|cc|cccc}
~& $\mathrm{Fe_3O_4/MgO}$ & $\mathrm{Fe_3O_4/NiO/MgO}$ & B-term. & $\mathrm{Fe_{1-\delta}}$ (NM) & $\mathrm{Fe_{1-\delta}}$ (FM)& Chang \\ 
\hline
$\Delta z_{708.4\,\mathrm{eV}}$ (\r{A})& 0 & 0 & 0& 1.7 & 0 & 0 \\ 
$\Delta z_{709.5\,\mathrm{eV}}$ (\r{A})& $3.2\pm 1.3$ & $2.6\pm 0.9$ & 2.5 & 4.1 & 5.5 & 2.5\\ 
$\Delta z_{710.2\,\mathrm{eV}}$ (\r{A})& 0 & $3.3\pm 1.4$& 0 & 1.7 & 1.1 & -0.5\\ 
\end{tabular}
\caption{\label{p3_tabdistances} Distances of the rising edges of the magnetooptical profiles from the structural interface at the three resonant energies. Considered here are the results from the two investigated samples as well as the discussed interface models: the B-terminated model (cf. Fig.~\ref{p3fig_simProf}), the non-magnetic (NM) 
$\mathrm{Fe_{1-\delta}}$ interface layer (cf. Figs.~\ref{p3_figProfSim_extra}(a)-(c)), the ferromagnetic (FM) $\mathrm{Fe_{1-\delta}}$ interface layer (cf. Figs.~\ref{p3_figProfSim_extra}(d)-(f)) and the model proposed by Chang et al. \cite{Chang16} (cf. Figs.~\ref{p3_figProfSim_extra}(g)-(i)).}
\end{table*}

\section{Discussion}
For the Fe\textsubscript{3}O\textsubscript{4}/MgO sample, it is shown in Fig. \ref{p3fig_expProf}(b) that the magnetooptical depth profile recorded at a photon energy of $709.5\,$eV is displaced from the interface into the Fe\textsubscript{3}O\textsubscript{4} film by a shift $\Delta z_\mathrm{709.5\,eV}$.
From the quality of the fits, we can determine this shift to the range $\Delta z_\mathrm{709.5\,eV}= 3.2\pm 1.3$\r{A}. 
Since this resonance is governed by the tetrahedrally coordinated Fe ions, this result is consistent with a B-terminated interface having octahedrally coordinated $\mathrm{Fe^{2+}}$ and $\mathrm{Fe^{3+}}$-cations in the Fe\textsubscript{3}O\textsubscript{4} interface layer.
Figure \ref{p3fig_simProf}(a) shows the ideal stacking order at a B-terminated $\mathrm{Fe_3O_4/MgO(001)}$ interface. The oxygen lattice of the substrate continues as the oxygen lattice of the film. In [001] direction, $\mathrm{Fe_3O_4}$ can be described as a stack of subsequent B layers consisting of $\mathrm{O^{2-}}$ anions as well as $\mathrm{Fe^{2+}_{oct}}$ and $\mathrm{Fe^{3+}_{oct}}$ cations, and A layers containing $\mathrm{Fe^{3+}_{tet}}$ cations with a distance of $1.05\,$\r{A} between them. The stacking of B- and A-layers is depicted as blue and red shaded areas, respectively, in Fig. \ref{p3fig_simProf}(a) with the interface layer being a B layer (B-termination).
Simulated cation depth profiles following this model are shown in Fig. \ref{p3fig_simProf}(b). The atomically sharp distributions are smeared out using an interface roughness of $\sigma = 3.5\,$\r{A} corresponding to the experimentally determined roughness of the $\mathrm{Fe_3O_4/MgO}$ interface. The rising edge of the $\mathrm{Fe^{3+}_{tet}}$ depth profile is shifted by $\Delta z_\mathrm{tet}=1.05\,$\r{A} from the interfaces of the $\mathrm{Fe^{2+}_{oct}}$ and $\mathrm{Fe^{3+}_{oct}}$ 
profiles into the Fe\textsubscript{3}O\textsubscript{4} film. Because of the overlap of the individual cation spectra (cf. Fig. \ref{p3fig_XMCD}(b)), the expected magnetooptical depth profiles at the different resonant energies do not follow this behavior exactly. 

Taking into account the magnetooptical contributions as derived from the multiplet calculations of each cation at each of the three energies, the expected magnetooptical depth profiles of a B-terminated $\mathrm{Fe_3O_4/MgO(001)}$ interface can be calculated. They are shown in Fig. \ref{p3fig_simProf}(c). The expected shift of the magnetooptical depth profile at $709.5\,$eV is $\Delta z_\mathrm{709.5\,eV}=2.5\,$\r{A}, consistent with the experimental result of $\Delta z_\mathrm{709.5\,eV}=3.2\pm 1.3\,$\r{A}.  
Therefore, the magnetooptical depth profiles indicate a B-terminated $\mathrm{Fe_3O_4/MgO(001)}$ interface with no interlayer, and evidently, also no magnetic dead layer.

In this scenario, the magnetic order of all three sublattices has bulk properties down to the interface. 
The simulations of the B-terminated interface also predict that the apparent roughness $\sigma_\mathrm{708.4\,eV}$ of the magnetooptical depth profiles at $708.4\,$eV appears to be about $0.5\,$\r{A} larger than the structural profile $\sigma_\mathrm{substrate}$. This offers an explanation for the slight mismatch of these two roughnesses observed in the experiment.
However, while the discrepancy between the model distance and the experimental distance is well within the error range, it is still substantial enough to make a discussion of alternative models worthwhile.

Both the $\mathrm{Fe_3O_4/MgO}$ and the $\mathrm{Fe_3O_4/NiO}$ interfaces have been studied by various methods. Spintronic devices require interfaces that are structurally, but especially also magnetically sharp. Therefore, focus has been laid on the possible presence of interlayers and intermixing at the interfaces.
In the case of $\mathrm{Fe_3O_4}$ directly grown on a substrate, the formation of FeO interlayers has been reported on both metal and oxide substrates \cite{Gota99,Karunamuni99,Schlueter11}, and particularly on MgO(001) for films deposited at room temperature \cite{Bertram12}. 
The possibility of a single atomic $\mathrm{Fe_{1-\delta}O}$ interlayer is discussed in the following. 
The corresponding magnetooptical depth profiles are calculated in Figs. \ref{p3_figProfSim_extra}(a)-(f) for two scenarios. 

Therefore, the first reasonable scenario is that the $\mathrm{Fe_{1-\delta}O}$ interlayer forms a magnetic dead layer at the interface, presented in Figs. \ref{p3_figProfSim_extra}(a)-(c). Figure \ref{p3_figProfSim_extra}(a) shows an illustration of the stacking order, Fig. \ref{p3_figProfSim_extra}(b) the cation depth profiles simulated with a roughness of $\sigma=3.5\,$\r{A}, and Fig. \ref{p3_figProfSim_extra}(c) the resulting magnetooptical depth profiles. $\mathrm{Fe_{1-\delta}O}$ is paramagnetic at room temperature. Due to the magnetically dead $\mathrm{Fe_{1-\delta}O}$ layer, the rising edges of the magnetooptical depth profiles at both 708.4$\,$eV and 710.2$\,$eV are shifted about 2$\,$\r{A} into the Fe\textsubscript{3}O\textsubscript{4} film compared to the structural interface (cf. Tab.~\ref{p3_tabdistances}), not consistent with the observed profiles in Fig.~\ref{p3fig_expProf}(b).

The second scenario assumes that the very thin FeO layer magnetically couples to $\mathrm{Fe_3O_4}$ and the magnetic order of its $\text{Fe}^{2+}_\text{oct}$ sublattice is extended into the FeO layer. The resulting magnetooptical depth profiles show non-monotonic behavior and differ even more from the observed ones (cf. Figs. \ref{p3_figProfSim_extra}(d)-(f) and Tab.~\ref{p3_tabdistances}).

Our data therefore do not indicate any magnetically dead layers, which had been considered to be the cause of the magnetization reduction in $\mathrm{Fe_3O_4}$ ultrathin films \cite{VanDerHeijden96,Zhou06,Gomes14,Chang16}, nor interlayers of ferromagnetic FeO. 
Since it has already been shown that the magnetooptical depth profiles can be explained without an interlayer, it is unlikely that a FeO interlayer of more than a single atomic layer is present. 

Another interesting model, which was proposed by Chang et al. for the growth dynamics of $\mathrm{Fe_3O_4}$ \cite{Chang16} shall also briefly be mentioned here. This model suggests the first interface B-layer to contain one additional $\mathrm{Fe^{3+}_{oct}}$ cation per unit formula [$(\mathrm{Fe^{2+}_{oct}})_2(\mathrm{Fe^{3+}_{oct}})_3\mathrm{O_8}$ instead of $(\mathrm{Fe^{2+}_{oct}})_2(\mathrm{Fe^{3+}_{oct}})_2\mathrm{O_8}$]. An illustration can be seen in Fig. \ref{p3_figProfSim_extra}(g).
Both the cation depth profiles and the simulated magnetooptical depth profiles resulting from this model, presented in Figs. \ref{p3_figProfSim_extra}(h),(i), hardly differ from the B-terminated interface shown in Figs. \ref{p3fig_simProf}(b),(c).
Although XRMR would in principle be an ideal method to test this model, the lacking spatial resolution in our experiments can neither confirm nor reject a faint phenomenon like an additional $\mathrm{Fe^{3+}_{oct}}$ cation in the interface layer.

For the Fe\textsubscript{3}O\textsubscript{4}/NiO/MgO(001) sample, a noteworthy finding is the dichroic signal of the NiO film.
Bulk NiO is an antiferromagnet at room temperature and should not show any circular dichroism. However, as demonstrated in Fig. \ref{p3fig_NiMCD}, we clearly observe magnetooptical effects in both TFY mode and in reflection at the Ni $L_{3}$ edge.
Interestingly, we can exclude the XMCD signal to stem from uncompensated surface spins, since the magnetooptical depth profiles clearly show a homogeneous magnetization of the entire film.
Size-effects of the magnetic properties of NiO, including ferromagnetic behaviour at room temperature, have been frequently reported before, mostly for NiO nanoparticles \cite{Kodama97,Winkler05,Li06,Ravikumar15,Rinaldi-Montes16}. For $\mathrm{Fe_3O_4/NiO}$ ultrathin films, a spin-flop coupling of NiO to the $\mathrm{Fe_3O_4(001)}$ interface has been reported \cite{Krug08}. In that case, the antiferromagnetic order of NiO aligns perpendicular to the magnetization of $\mathrm{Fe_3O_4}$, but with a small canting of the $\mathrm{Ni^{2+}}$ moments, resulting in a magnetization component parallel to the ferrimagnet. This reaction of NiO to outer magnetic fields has also been confirmed by spin Hall magnetoresistance measurements \cite{Hoogeboom17,Fischer18}, and can explain the presence of the observed XMCD signal. 

The interface of $\mathrm{Fe_3O_4/NiO}$ has mostly been discussed regarding the presence of a $\mathrm{NiFe_2O_4}$ interlayer.
In reports by Gatel et al. \cite{Gatel05} and Pilard et al. \cite{Pilard07}, high-resolution transmission electron microscopy (HRTEM) images show generally sharp interfaces between the rock salt structure of NiO and the spinel structure of $\mathrm{Fe_3O_4}$. However, this only partly solves the question, because the lack of chemical and magnetic contrast cannot exclude, for instance, the formation of a rock salt $\mathrm{(Fe,Ni)O}$ phase or a spinel $\mathrm{NiFe_2O_4}$ phase. Gatel et al. adressed this issue by performing both HRTEM and electron energy loss spectroscopy (EELS) \cite{Gatel05}. For a 
$\mathrm{NiO/Fe_3O_4/MgO(001)}$ sample, for which the NiO film was deposited at a substrate temperature of $700^\circ$C, indeed an intermediate $\mathrm{NiFe_2O_4}$ phase was observed. For a $\mathrm{Fe_3O_4/NiO/MgO(001)}$ stack, which had the Fe\textsubscript{3}O\textsubscript{4} film deposited at $400^\circ$C, the interface appeared to be chemically sharp with at most minor interdiffusion \cite{Gatel05}.
The $\mathrm{NiFe_2O_4}$ phase they observe in the $\mathrm{NiO/Fe_3O_4/MgO(001)}$ stack is likely caused by thermal interdiffusion due to the high deposition temperature \cite{Kuschel16}. 

A formation of a well-ordered $\mathrm{NiFe_2O_4}$ interlayer does not match our observations. $\mathrm{NiFe_2O_4}$ crystallizes in the same inverse spinel structure as Fe\textsubscript{3}O\textsubscript{4}, but with $\mathrm{Ni_{oct}^{2+}}$ cations instead of $\mathrm{Fe_{oct}^{2+}}$ sharing the B-sites with $\text{Fe}^{3+}_\text{oct}$. Analogous to Fe\textsubscript{3}O\textsubscript{4}, the magnetic moments of the $\mathrm{Ni_{oct}^{2+}}$ and $\text{Fe}^{3+}_\text{oct}$ cations on the B-sites align antiferromagnetically to the magnetic moments of the $\text{Fe}^{3+}_\text{tet}$ cations on the A-sites. 
For the magnetooptical depth profiles, this would imply a decrease of $\mathrm{Fe^{2+}_{oct}}$ cations close to the interface and thus of the magnetooptical absorption at $708.4\,$eV, while the magnetooptical absorption depth profiles for $709.5\,$eV and $710.2\,$eV should stay constant in a $\mathrm{NiFe_2O_4}$ layer. Instead, we observe a reduction of the magnetooptical absorption at $709.5\,$eV and $710.2\,$eV as compared to the case of the $\mathrm{Fe_3O_4/MgO}$ interface. 
This behavior, together with the increased roughnesses of the magnetooptical depth profiles at the $\mathrm{Fe^{2+}_{oct}}$ and the $\mathrm{Ni^{2+}}$ energies, might indicate a slight interdiffusion of $\mathrm{Fe^{2+}_{oct}}$ into the rock salt structure of $\mathrm{NiO}$. However, this effect may not extend farther than a single atomic layer. Notably, the ferromagnetic coupling between the $\text{Fe}^{2+}_\text{oct}$ cations is retained down to the interface regardless of the intermixing.

\section{\label{sec:level6}Conclusion}
We have prepared ultrathin $\mathrm{Fe_3O_4/MgO(001)}$ and $\mathrm{Fe_3O_4/NiO/MgO(001)}$ films by RMBE and performed XMCD and XRMR measurements to extract magnetooptical depth profiles for the individual cation species $\mathrm{Fe^{2+}_{oct}}$, $\mathrm{Fe^{3+}_{tet}}$ and $\mathrm{Fe^{3+}_{oct}}$ as well as for $\mathrm{Ni^{2+}_{oct}}$. These magnetooptical depth profiles show that for $\mathrm{Fe_3O_4/MgO(001)}$, the magnetic order of all three cation species is stable for the entire film with no interlayer or magnetic dead layer at the interface. For $\mathrm{Fe_3O_4/NiO}$ films, we observe a magnetooptical absorption at the Ni $L_3$ edge in the NiO film corresponding to uncompensated magnetic moments throughout the entire NiO film.
The magnetooptical profiles of the iron cations reveal an intact magnetic order for the $\text{Fe}^{2+}_\text{oct}$ cation species down to the interface, while the magnetooptical depth profiles at the $\mathrm{Fe^{3+}_{oct}}$ and the $\mathrm{Fe^{3+}_{tet}}$ resonances are shifted about $3\,$\r{A} into the Fe\textsubscript{3}O\textsubscript{4} film, possibly indicating a single intermixed layer containing both $\mathrm{Fe^{2+}}$ and $\mathrm{Ni^{2+}}$ cations. 
\section{\label{sec:level7} Acknowledgements}
Financial support from the Bundesministerium f\"ur Bildung und Forschung (FKZ 05K16MP1) is gratefully acknowledged.  We are also grateful for the kind support from the Deutsche Forschungsgemeinschaft (DFG under  No. KU2321/6-1, and No. WO533/20-1).
The authors would like to thank Diamond Light Source for beamtime (proposal SI19173-1), and the staff of beamline I10 for assistance with data collection.
We thank HZB for the allocation of synchrotron beamtime at beamline $\mathrm{UE46\_PGM}$-1 $\text{(181/06266ST/R)}$ where we recorded the XRMR and XMCD measurements.
\bibliography{FeONiOBib}
\end{document}